\documentclass[twoside]{article}

\usepackage{graphicx}
\usepackage{amsmath}
\usepackage{amssymb}
\usepackage{bm}

\usepackage[T2A]{fontenc}
\usepackage[cp1251]{inputenc}
\usepackage{cmpj2e}

\title
{Revealing Novel Quantum Phases in Quantum Antiferromagnets on Random Lattices}%

\author
{Rong Yu\refaddr{label1,label2}, Stephan Haas\refaddr{label4}, and
Tommaso Roscilde\refaddr{label3}}
\addresses{
\addr{label1} Department of Physics and Astronomy, University of
Tennessee, Knoxville, TN 37996, USA \addr{label2} Material Science
and Technology Division, Oak Ridge National Laboratory, Oak Ridge,
TN 32831, USA
\addr{label4} Department of Physics and Astronomy, University of
Southern California, Los Angeles, CA 90089-0484, USA
\addr{label3} Laboratoire de Physique, \'Ecole Normale
Sup\'erieure de Lyon, 46 All\'ee d'Italie, 69003 Lyon, France}
\begin{document}

\maketitle

\begin{abstract}
Quantum magnets represent an ideal playground for the controlled realization of
novel quantum phases and of quantum phase transitions. The Hamiltonian
of the system can be indeed manipulated by applying a magnetic field
or pressure on the sample. When doping the system
with non-magnetic impurities, novel inhomogeneous phases emerge
from the interplay between geometric randomness and quantum fluctuations.
In this paper we review our recent work on quantum phase transitions and
novel quantum phases realized in disordered quantum magnets.
The system inhomogeneity is found to strongly affect phase transitions by
changing their universality class, giving the transition a novel, quantum
percolative nature. Such transitions connect conventionally ordered phases
to unconventional, quantum disordered ones - quantum Griffiths phases,
magnetic Bose glass phases - exhibiting gapless spectra associated with
low-energy localized excitations.

\pacs 
\end{abstract}

\section{Introduction}

Quantum phase transitions (QPTs) and related collective quantum phases
represent one of the most exciting research topics in
condensed matter physics \cite{Sondhietal97}. Contrary to thermal phase transitions,
QPTs occur at zero temperature upon tuning a parameter of the system Hamiltonian.
The emergence of quantum collective phenomena opens the path towards
quantum phases that do not admit any classical counterpart. Quantum magnets
provide a large showcase of materials in which QPTs have been experimentally
demonstrated. Quantum fluctuations, driving the
system through a QPT, can be continuously tuned by \emph{e.g.} applying
a magnetic field or by exerting pressure on the sample to control the
magnetic couplings among the spins.
 A well-known example is represented by magnetic Bose-Einstein
 condensation (BEC) in spin-gap materials~\cite{Giamarchi08},
 such as systems of weakly coupled $S=1/2$ dimers.
 These magnetic insulators show a paradigmatic quantum-disordered
ground state, namely a total singlet state with a gap to all triplet
excitations. Application of a magnetic field can close the triplet
gap, inducing the appearance of bosonic spin triplets in the
ground state of the system, forming a Bose condensate of magnetic
quasiparticles. This condensed state corresponds to a magnetically
ordered state with spontaneous appearance of a staggered magnetization
transverse to the field.

\begin{figure}[htb]
\centerline{\includegraphics[width=0.8\textwidth]{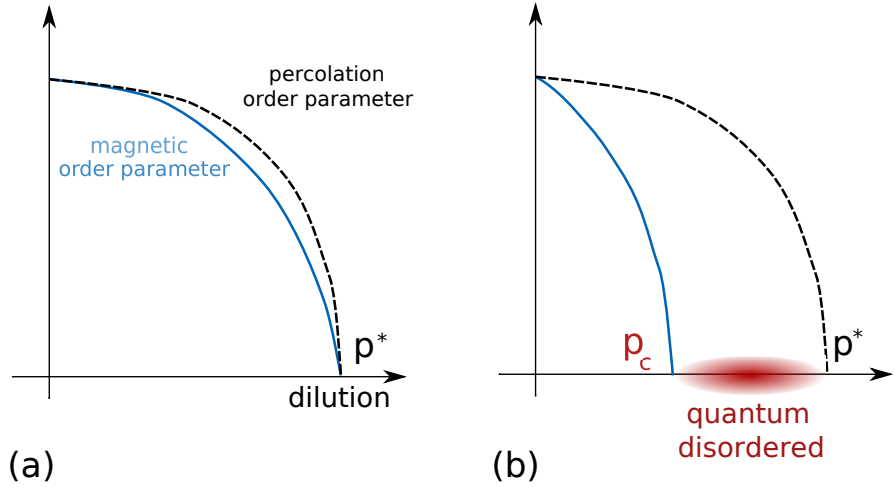}}
\caption{(a) Conventional percolation transition in a magnet: upon diluting the system
the magnetic order parameter is found to vanish at the percolation threshold $p^*$ .
(b) Disorder-induced quantum phase transition: at a critical dilution $p_c < p^*$
the magnetic order disappears, leaving space to a novel quantum disordered
phase.}
\label{f.qperc}
\end{figure}

 A completely different route towards QPT transitions in quantum
 magnets is represented by the effect of non-magnetic doping.
 Starting from a magnetically ordered state, a simple route towards
 disordering the system is by diluting the lattice via site or bond
 removal. A genuine quantum phase transition is realized
 if quantum fluctuations lead to the loss of magnetic order
 at $T=0$ before the lattice reaches the percolation threshold.
 Experimental studies on model magnets, such
 as non-magnetically doped La$_2$CuO$_4$ \cite{Vajketal02},
 supported by extensive numerical studies \cite{Sandvik02}
 show that fundamental quantum models like the
 two-dimensional quantum Heisenberg antiferromagnet
 do not develop strong enough quantum fluctuations for
 a quantum phase transition to occur. Yet, increasing quantum
 fluctuations \emph{ad hoc}, \emph{e.g.} by explicit lattice dimerization
 \cite{VajkG02,Sandvik02B,Sknepneketal04} or by anisotropic bond dilution
 \cite{Yuetal05}, can lead indeed to disorder-induced
 quantum phase transitions (see Fig.\ref{f.qperc}). And what better strategy could
 one envision to tune quantum fluctuations than to consider
 a system which already exhibits a quantum phase transition
 in the clean limit?

   This idea brings naturally to the investigation of the effect
 of disorder on quantum critical phenomena in magnetic systems,
 and in particular on magnetic BEC as a paradigmatic
 magnetic quantum phase transition
 \cite{RoscildeH05,RoscildeH06,Roscilde06,Yuetal08}.  Introducing disorder
 in the magnetically ordered phase immediately close to
 the condensation QPT corresponds to exposing
 the dilute gas of spin triplets induced by the field to a
 random potential: as in any system of weakly interacting
 quantum particles, this leads to Anderson localization
 for moderate (or even infinitesimal) disorder strengths \cite{LeeR85}.
 This property translates into the possibility of disrupting
 the field-induced magnetic order by diluting the lattice
 well below the percolation threshold, accomplishing
 in this way a genuine disorder-induced quantum phase
 transition.

\begin{figure}[htb]
\centerline{\includegraphics[width=0.4\textwidth, angle=270]{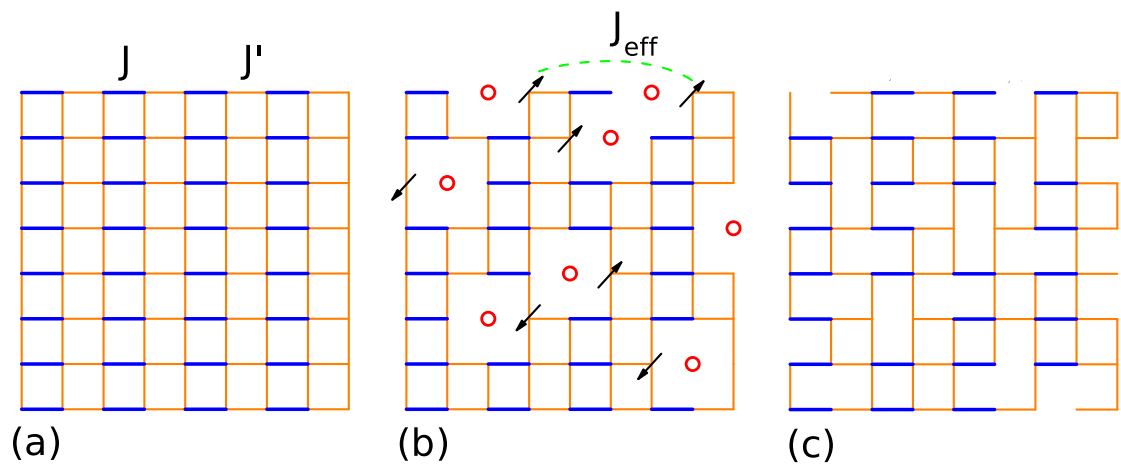}}
\caption{(a) A square lattice decomposed into dimer and inter-dimer bonds.
The dimer bonds are shown in dark color with coupling $J$, and the
inter-dimer bonds are shown in light color with coupling $J'$. Both
$J$ and $J'$ are antiferromagnetic. (b) Site dilution of the dimer
singlet ground state releases random magnetic local moments which
are coupled through long-range coupling
$J_{\rm{eff}}$. The circles refer to the non-magnetic impurities,
and arrows correspond to local moments near the impurity sites.
(c) Inhomogeneous bond dilution of the same lattice: the dimer bonds ($J$)
are populated with probability $P$, while the inter-dimer bonds ($J'$) are populated
with probability $P'$.}
\label{f.dimers}
\end{figure}

  A side effect of doping weakly coupled dimer systems
 is the appearance of unpaired $S=1/2$ local moments,
 which are coupled through virtual excitations of the
 intact dimers \cite{SigristF96,Roscilde06,Mikeskaetal04} (see Fig.\ref{f.dimers}(a)).
 The resulting random network of
 interacting local moments can support long-range
 antiferromagnetic order in zero field, giving rise to a counter-intuitive
 phenomenon of order by disorder (OBD).
 This ordered state is found to be disrupted
 by the application of a moderate magnetic field \cite{Roscilde06,Yuetal08}.
 The resulting disordered phase shows
 short-range antiferromagnetic correlations surviving
 on localized regions, which can be mapped onto
 Anderson localized quasiparticles.

 Hence we see in general that disorder-induced
 quantum phase transitions drive the system to novel quantum
disordered phases with rather exotic properties. A common
denominator of these phases is a gapless spectrum associated
with the appearance of low-energy localized excitations,
associated with exponentially rare regions in the
disordered lattice. If the energy cost of the localized
excitation decreases algebraically with the inverse size of the region
that hosts it, all conventional response functions remain
non-singular in this phase down to $T=0$. On the other
hand, if the same energy cost decreases exponentially
when the size of the region increases, this leads to quantum
Griffiths singularities \cite{Griffiths} with a non-universal, disorder-dependent
divergent behavior in measurable quantities such as
the uniform susceptibility.

 In this paper we review our recent progress in studying
 disorder-induced QPTs as well as the effect of disorder
 on field-induced QPTs in two-dimensional quantum magnets.
This topic is particularly challenging from a technical point of
view. Indeed conventional perturbation methods are generally
doomed to fail at quantum critical points;  renormalization group
schemes, while generally successful for the study of critical points
in clean systems, prove to be particularly hard for critical
points in presence of disorder \cite{Weichman08}.
A third way is provided by numerically exact methods,
which are particularly well developed for unfrustrated
quantum spin systems. In particular we make use of
the stochastic series expansion quantum Monte Carlo
method \cite{SyljuasenSandvik02} which proves to be a very
powerful quantum Monte Carlo technique to investigate the
low-temperature properties of quantum spin systems on a large scale.
The non-local nature of the update algorithm makes it possible
to circumvent critical slowing down and to investigate quantum
critical phenomena in a very accurate
way.  While disorder imposes the extra cost of averaging over
disorder statistics, this task can be accomplished on modern
supercomputers.

The structure of the paper is as follows. In
section~\ref{s.bond} we show how the interplay between anisotropic
geometric randomness and quantum fluctuations gives rise to a new class
of percolative quantum phase transitions in an inhomogeneous
bond-diluted antiferromagnet in two dimensions (2D). This transition
opens a novel quantum disordered phase which exhibits quantum
Griffiths singularities.
In section~\ref{s.BG}, we analyze the very rich phase diagram
of site-diluted weakly coupled dimer systems in a magnetic
field, showing the occurrence of and extended magnetic Bose glass
phase.

\section{Percolative Quantum Phase Transition in Strongly Fluctuating Quantum
Antiferromagnets} \label{s.bond}

 Our starting point is the quantum Heisenberg antiferromagnetic model defined
on a regular bipartite lattice, as illustrated in Fig.~\ref{f.dimers}(a), with
Hamiltonian
\begin{equation}
{\cal H} =  \sum_{\langle ij \rangle} J_{ij} {\bm S}_i \cdot {\bm S}_j +
 \sum_{\langle lm \rangle} J'_{lm} {\bm S}_l \cdot {\bm S}_m - g\mu_b H \sum_i S_i^z
\end{equation}
where $\bm S$ are $S=1/2$ spin operators. In the most general case,
two sets of bonds with different local strengths ($J_{ij}$ and $J'_{lm}$)
have been singled out. In the clean case one has $J_{ij} = J$ for all $\langle ij \rangle$
and $J'_{lm} = J'$ for all $\langle lm \rangle$. In the following we will
focus our attention on 1) the case of a planar array of dimers in a square
lattice (as specifically illustrated by Fig.~\ref{f.dimers}(a)) and
2) the case of a bilayer system, in which the bonds of strength $J'$
form two square lattices which are connected by the bonds of strength $J$.
In the case 1) of a simple square lattice, the choice $J = J'$
reproduces the well-known limit of the two-dimensional quantum Heisenberg
antiferromagnet (2DQHAF). In this section we will consider the case of
zero applied magnetic field $H$, while the effects of a finite field will be
accounted for in the next section.

 In the clean limit, the 2DQHAF is well known to support long-range
N\'eel order\cite{Manousakis91}. As mentioned in the introduction, a central question of
quantum magnetism is the response of magnetically ordered systems
to the dilution of their magnetic lattice, either in the form of bond or site
dilution. From a geometric point of view, bond
or site dilution reduce the connectivity of the lattice, ultimately
leading to a percolative phase transition~\cite{StaufferA94} beyond
which the system is broken up into finite clusters. In a classical
spin system, this percolation transition is coupled to a magnetic
transition with the same critical exponents, since spontaneous
magnetic order cannot survive beyond the percolation threshold. In a
quantum spin system, on the other hand, a progressive reduction of
the lattice connectivity enhances quantum fluctuations in a
continuous fashion, raising the possibility of quantum destruction
of magnetic order {\em before} the percolation threshold is reached.
Recently, the evolution of the magnetic state of the 2DQHAF under
site or bond dilution has been studied extensively both in experiment and theory.
Experimentally, the effect of site dilution has been probed in
La$_2$Cu$_{1-p}$(Zn,Mg)$_p$O$_4$, in which magnetic
Cu$^{2+}$ ions are replaced randomly by non-magnetic Zn$^{2+}$ or
Mg$^{2+}$ ions~\cite{Vajketal02}. The fundamental result of this
study is that magnetic order at low temperature disappears
at a critical dilution $p=p_c$ which coincides with the
percolation threshold $p^{*}=0.40725$~\cite{StaufferA94}.
 This has been further confirmed by extensive numerical
simulations both in the case of site and bond dilution~\cite{Sandvik02},
showing that the percolating cluster supports long-range order up to
the percolation threshold; this means that magnetism can only be
discarded geometrically by fragmenting the percolating cluster beyond
$p^*$.

\begin{figure}[htb]
\centerline{\includegraphics[width=0.65\textwidth]{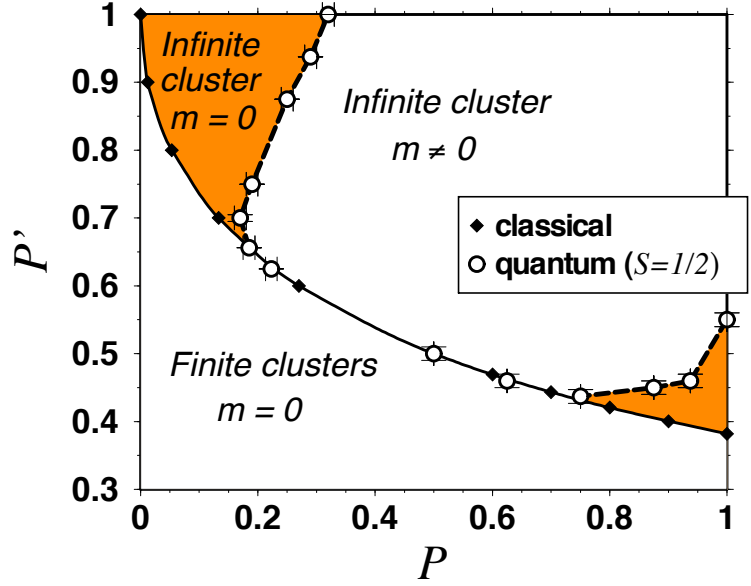}}
\caption{Phase diagram of the
inhomogeneously bond-diluted $S=1/2$ antiferromagnet on the square
lattice. The colored area indicates the quantum-disordered region in
which the system has developed an infinite percolating cluster, but
the magnetization $m$ of the system vanishes because of quantum
fluctuations.}
\label{f.PhDBondDilute}
\end{figure}

An alternative scenario to the above classical percolation picture
is offered by \emph{quantum percolation}, in which the geometric
transition and the magnetic one are decoupled by quantum
fluctuations. This scenario invokes the fact that spins involved in locally
strongly fluctuating quantum states, such as dimer singlets and
resonating valence bond states, are weakly correlated with the
remainder of the system. In a random network of spins, the local
strongly fluctuating states create \emph{weak links} with small
spin-spin correlations. If these weak links are part of the backbone
of the percolating cluster, they can prevent the percolating cluster
from developing long-range order. Therefore, if lattice dilution
favors the local formation of such states, it is possible to drive
the system towards a quantum disordered state before the percolation
threshold is reached, decoupling percolation from
magnetic ordering.

This quantum percolation scenario has been recently
demonstrated for the $S=1/2$ Heisenberg model on the
bilayer lattice under dimer dilution~\cite{Sandvik02B,VajkG02,Sknepneketal04}.
In this model quantum fluctuations can be arbitrarily tuned
by increasing the strength of the inter-layer coupling $J$ with
respect to the intra-layer one $J'$: indeed, even in the clean
model, for $J> J_c = 2.5 J'$ the spins on each inter-layer dimer form
a singlet and long-range order is lost \cite{SandvikS94}. Randomly diluting the bilayer
by taking away a percentage $p$ of \emph{whole dimers} at once
leads to the percolation of the bilayer lattice for $p=p^*$,
analogously to what happens on a simple square lattice.
For $J \lesssim J^* =   0.16 J'$ the bilayer shows long-range
order up to the percolation threshold \cite{Sandvik02B}.
The tuning knob of quantum fluctuations offered by the
inter-layer coupling allows to destroy long-range order
on the percolating cluster for $J \gtrsim J^*$: beyond this
value, geometric percolation of the lattice is no longer
a sufficient condition for long-range magnetic ordering,
due to the extremely strong quantum fluctuations on
dimers with lower local connectivity. The quantum disordered
phase appearing on a percolated lattice of dimers still lacks
a complete characterization, but it is legitimate to suspect
that, for $J < J_c$, rare but arbitrarily large regions which
are devoid of dimer vacancies can support locally
magnon-like excitations similar to those of the clean
system in its magnetically ordered phase. Hence the
spectrum of the system is expected to be \emph{gapless},
but not leading to a singular contribution to the response
functions (as shown by the regular behavior of the
uniform susceptibility in Ref.~\cite{Sandvik02B}).
A very similar picture of quantum percolation has been found by two
of us in the anisotropic $S=1$ Heisenberg antiferromagnet with
site dilution \cite{RoscildeH07}: here the quantum disordered phase has been identified
with a \emph{Mott glass}, namely a phase with a gapless spectrum and
a vanishing compressibility.

 In Ref.~\cite{Yuetal05} we have shown that a similar quantum
percolation scenario can be achieved in the \emph{standard} 2DQHAF
under \emph{inhomogeneous} bond dilution. In our model
all bonds are of equal strength $J=J'$, and quantum fluctuations
are instead enhanced in a purely \emph{geometrical} fashion by
the lattice randomness.  As illustrated in Fig.~\ref{f.dimers}(a), a square lattice can be
geometrically decomposed into dimers and ladders (made of inter-dimer
bonds) in such a way that there are no two adjacent dimers or
ladders. The inhomogeneous bond dilution is realized by assigning
different occupation probabilities to intra-dimer and inter-dimer
bonds. Explicitly

\begin{equation}
\text{dimer bonds}~~ J_{ij} = \begin{cases}
~J ~~~~~ \text{with probability}~~ P \\
~0 ~~~~~ \text{with probability}~~ 1-P
\end{cases}
\end{equation}

\begin{equation}
\text{ladder bonds}~~ J'_{lm} = \begin{cases}
~J ~~~~~ \text{with probability}~~ P' \\
~0 ~~~~~ \text{with probability}~~ 1-P'~~~.
\end{cases}
\end{equation}

 As it is well-known, $S=1/2$ spin ladders have a quantum disordered
 singlet ground state \cite{DagottoR96}, and the same applies to weakly
 connected dimer lattices, such as the comb lattice. This special
 property of the $S=1/2$ Heisenberg model on low-dimensional
 lattices is the key of our dilution scheme. In fact, \emph{inhomogeneous}
 bond dilution $P \neq P'$
 (see Fig.\ref{f.dimers}(c)) favors the appearance of ladder segments
 ($P<P'$) or of weakly connected dimers ($P>P'$), both of which have the
 tendency to support locally a strongly fluctuating quantum state with
a significant singlet component, and effectively decoupled
from the rest of the percolating cluster.
 When such geometrical structures appear on the backbone
of the percolating cluster, they can effectively break it into
sub-clusters from the point of view of magnetic correlations,
leading to a quantum-disordered ground state.

The phase diagram of this inhomogeneous bond dilution model, as
resulting from an extensive classical and quantum Monte Carlo study \cite{Yuetal05},
is presented in Fig.~\ref{f.PhDBondDilute}. When the inhomogeneity
is weak ($P\approx P'$), the magnetic transition is found to coincide
with the percolation one, both in terms of the location of the transition
and in terms of its critical exponents. This is in agreement with the
findings in the homogeneous limit
$P=P'$ already investigated in Ref.~\cite{Sandvik02}. However,
for strong enough inhomogeneity the magnetic transition deviates
from the geometric one, turning into a quantum phase
transition beyond two multi-critical points. The critical
exponents, extracted from a finite-size scaling analysis of the correlation length,
confirm this crossover from \emph{classical} to \emph{quantum} percolation:
$\nu=4/3$ and $z=1.9\approx D=91/48$ (the fractal dimensions
of the percolation cluster at threshold \cite{StaufferA94}) are found for weak to
intermediate inhomogeneity, while $\nu=1$ and $z=1$ are found for sufficiently
strong inhomogeneity~\cite{Yuetal05}.

\begin{figure}[htb]
\includegraphics
[width=5.5cm]{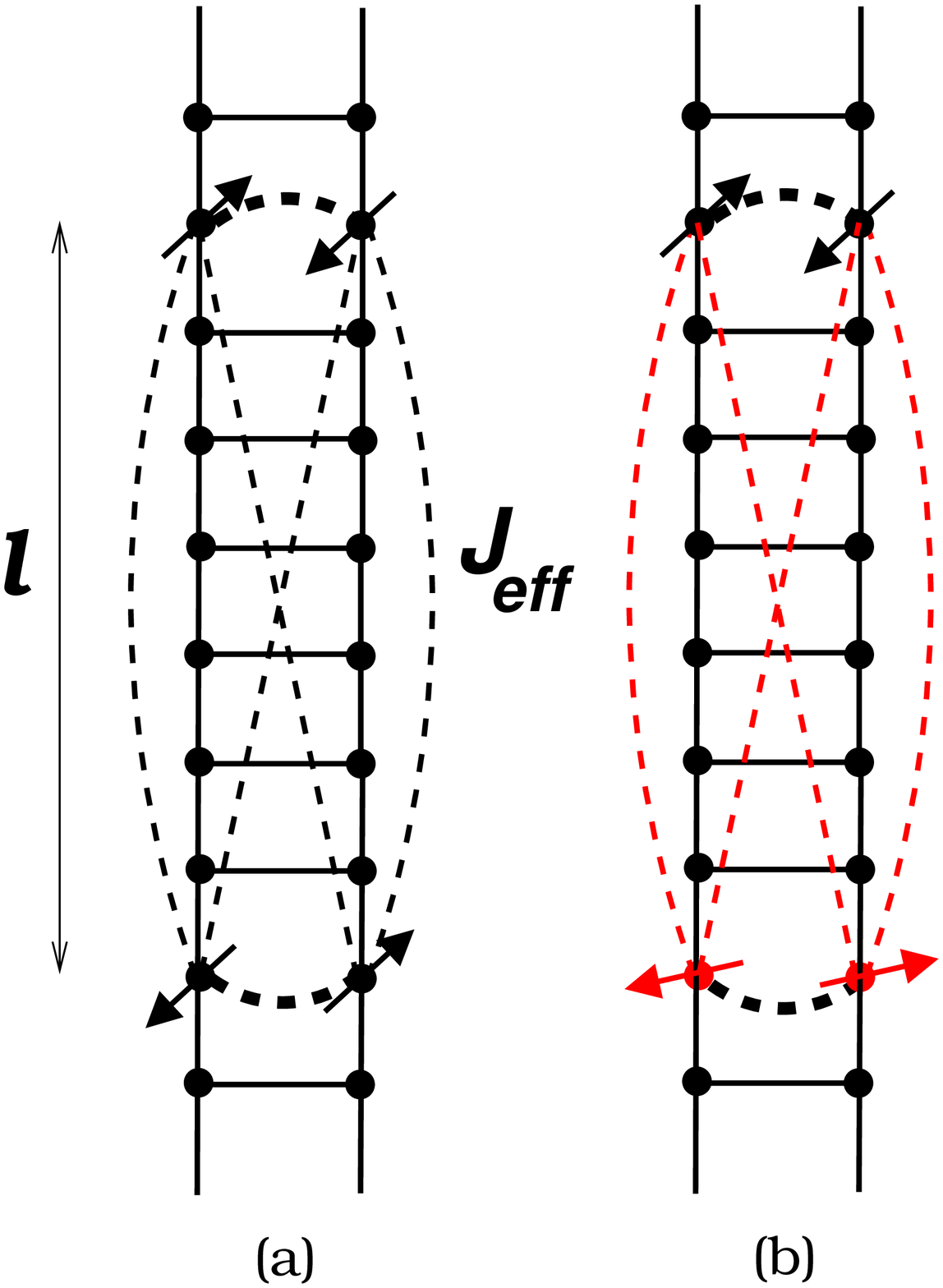}~~~~
\includegraphics[width=8.5cm]{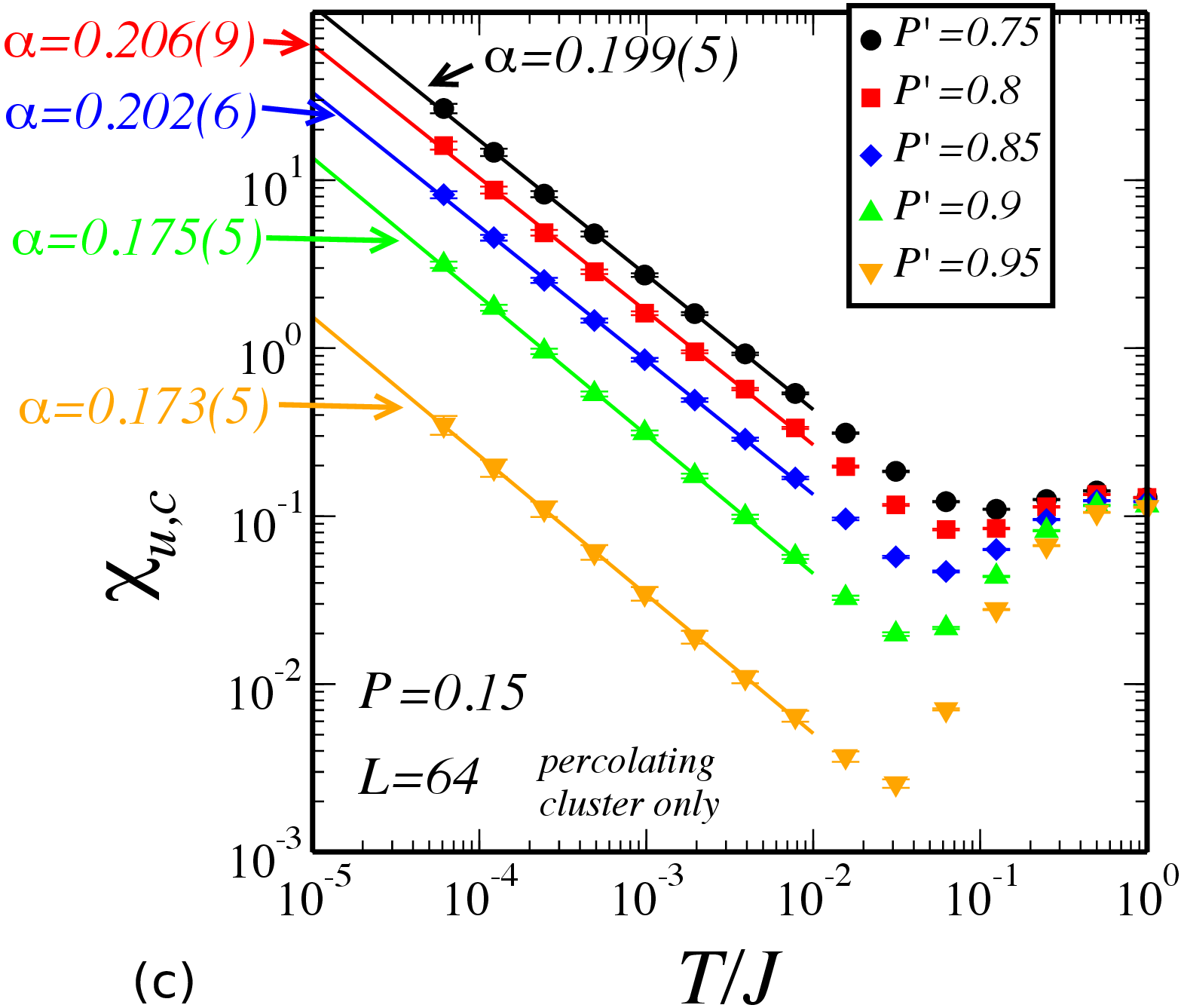}
\caption{(a)-(b) Sketch of a low-energy excitation in the
bond-diluted ladder system. (c)
Low-temperature uniform susceptibility of the percolating cluster only,
 $\chi_{u,c}$, in the quantum-disordered (ladder-like) regime. The solid lines are
power-law fits of the form $\chi_{u,c}\sim T^{-1+\alpha}$, and the
resulting fit coefficients $\alpha$ are indicated.  The percolating
cluster is picked as the largest cluster in a $64\times 64$ lattice.
The non-universal $\alpha$ values indicate the existence of
Griffiths singularities in this quantum-disordered phase.}
\label{f.chiuTGriffiths}
\end{figure}

As seen in Fig.~\ref{f.PhDBondDilute}, the bifurcation of the magnetic
transition line with respect to the percolation transition one
at strong enough inhomogeneity opens up two quantum disordered phases,
characterized by the existence of an infinite percolating clusters but zero
staggered magnetization $m$. Real-space spin-spin correlations in this phase are
short-ranged ~\cite{Yuetal06}. But the perfectly correlated
disorder in imaginary time  allows for
long-range correlations in this extra dimension, giving
rise to so-called quantum Griffiths singularities \cite{Griffiths}. As already
mentioned in the introduction, the dilution of the
lattice introduces local $S=1/2$ moments. If one
dilutes a lattice developing a spin-gapped singlet
ground state with a finite correlation length $\xi_0$
(as it will be the case in the following section), or
if dilution induces the local formation of such
singlet states (as it is the case in this section),
the resulting local moments develop mutual
effective interactions via the exchange of
virtual gapped excitations of the intermediate
regions. The effective couplings, $J_{\rm eff}$,
decay exponentially with the distance,
as a result of the finite ``mass" of the
exchanged excitation.
For two sites $i$ and $j$ with positions $\bm r_i$, $\bm r_j$, we
have that \cite{SigristF96,Roscilde06,Mikeskaetal04}
\begin{equation}
J_{\rm eff}(i,j) \sim J (-1)^{\bm r_i - \bm r_j} \exp(-|\bm r_i - \bm r_j|/\xi_0)
\label{e.Jeff}
\end{equation}
It is important to notice that the staggering prefactor eliminates
any frustration effect.
 At variance with site dilution (compare Fig.\ref{f.dimers}(b)), for bond
 dilution these local $S=1/2$ moments always occur in pairs.
 The effective interaction between two closeby moments is
 the strongest ($\sim J$), so that the lowest-energy excitations are not
 associated with exciting a pair of neighboring moments, but
 rather exciting two pairs of such moments lying far apart from
 each other by rotating one pair with respect to the other
 (see Fig.~\ref{f.chiuTGriffiths}(a)-(b)). According to Eq.~\ref{e.Jeff},
 this excitation has an energy
 scaling exponentially to zero with the inter-pair distance $l$, but
 the probability of existence of two pairs separated by a large
 distance is conditioned by the presence of a clean region
 in between them, whose probability in a diluted system is
 also exponentially suppressed with its size. In the case depicted
 in Fig.~\ref{f.chiuTGriffiths}(a)-(b), this probability is $(P')^{nl} = \exp(-n|\ln P'| l)$
 (where $n$ is the characteristic number of bonds present in a region of
 linear size $l$). The presence of an exponentially rare local excitation
 with exponentially small energy leads to a cancellation of the
 two exponentials in the calculation of the contribution of such
 excitations to fundamental response functions, as, \emph{e.g.},
 the uniform susceptibility. The result is a paradigmatic
 quantum Griffiths effect \cite{Yuetal06}.
 Indeed the uniform susceptibility of the percolating cluster
 in the quantum disordered phase displays a
non-universal power-law divergence when $T\rightarrow0$, as
presented in Fig.~\ref{f.chiuTGriffiths}(c). It is important to
stress that here we only consider percolating clusters with an
\emph{even} number of particles, which means that the ground state
on a finite-size system is expected to be a total spin singlet.
Hence the presence of a non-universal divergence of the
uniform susceptibility is not due to the presence of paramagnetic
dangling spins, but it is a genuine result of the fact that the
spectrum over the singlet ground state is gapless.

 From an experimental point of view, bond dilution can be
 considered as a limit of bond disorder introduced by
 doping the non-magnetic atoms/ions lying on the
 super-exchange paths responsible for the occurrence
 of antiferromagnetic couplings. Hence inhomogeneous bond
 disorder can be introduced in antiferromagnets in which
 super-exchange paths are mediated by different atomic
 species along different spatial directions. This is
 the case of many spin-ladder compounds, such as \emph{e.g.}
 (C$_5$H$_{12}$N$_2$)$_2$CuBr$_4$, in which
 intra-ladder and inter-ladder couplings between Cu$^{2+}$
 ions are mediated by chemically different non-magnetic ions.
 From a more fundamental point of view, our results,
 along with those on dimer-diluted lattices \cite{VajkG02,Sandvik02B,Sknepneketal04}
 and on diluted anisotropic $S=1$ models \cite{RoscildeH07}
 show that disorder-induced magnetic transitions can be
 pushed arbitrarily far from percolation thresholds.

\section{Field-Induced Quantum Disordered States in Site-Diluted
Lattices}\label{s.BG}

In the previous section, we have seen that lattice dilution can
locally enhance quantum fluctuations in a quantum magnet,
possibly discarding long-range order. In this section the critical
role of quantum fluctuations will be tuned arbitrarily by driving the
system close through a quantum phase transition in presence of
disorder.

As already mentioned in the introduction, the quantum phase
transition in question is represented by magnetic Bose-Einstein condensation
in a system of weakly coupled $S=1/2$ dimers in a magnetic
field. As in the previous section, we consider two different geometries leading to
essentially the same physics: a bilayer geometry with interlayer
couplings $J$ and intra-layer couplings $J'$; a planar dimer array
(compare Fig.~\ref{f.dimers}(a)), with dimer couplings $J$
and inter-dimer couplings $J'$. In absence of
lattice dilution, for both geometries a ratio of couplings
$J/J'$ overcoming a critical value
$(J/J')_c$ ( $\approx 2.5$ for the bilayer system,
$\approx 1.91$ for the planar dimer system \cite{Matsumotoetal01}) stabilizes a quantum
disordered ground state with dimer singlets on the strongest
magnetic bonds. In both systems, the application of a magnetic field
$h = g\mu_b H/J$ (in reduced units) leads to condensation of triplet
excitations  at a lower critical field $h_{c1}^{(0)}$ and to the
appearance of staggered magnetic order transverse to it
~\cite{Giamarchi08}. Increasing the field up to an upper
critical field $h_{c2}^{(0)}$ leads to full polarization of the
spins and to the destruction of spontaneous order.

\begin{figure}[htb]
\centerline{\includegraphics[width=0.65\textwidth]{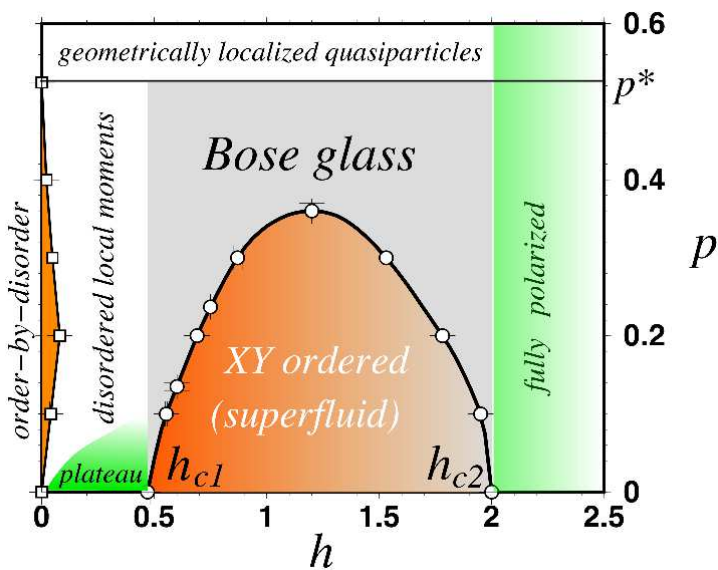}}
\caption{Ground-state phase diagram of the site-diluted bilayer
Heisenberg antiferromagnet with $J/J'=4$ in the field-dilution
plane. Ordered phases are indicated in orange, gapped disordered
phases are indicated in green, and gapless disordered phases in
white and grey.} \label{f.PhDBG}
\end{figure}

Doping the system induces a very rich phase diagram at $T=0$, as illustrated
in Fig.~\ref{f.PhDBG}. If the precise location of the phase boundaries
depends on the model parameter $J/J'$ as well as on the lattice
geometry, the qualitative features of this phase diagram are actually
independent of the value of $J/J'$, as long as $J/J' > (J/J')_c$,
and of the specific geometry of the dimer array (either bilayer
or planar array or other unfrustrated lattices).
For the specific case shown in Fig.~\ref{f.PhDBG} (bilayer system with $J/J'=4$),
the clean limit $p=0$ features Bose-Einstein condensation
transitions at $h_{c1}^{(0)}\approx 0.5 $  and $h_{c2}^{(0)}=2$.
Introducing site dilution of the bilayer lattice leads to an
upward shift of the lower critical field $h_{c1}$ with respect
to its clean value $h_{c1}^{(0)}$, and a downward shift
of the upper critical field. This means that diluting the system
at fixed field starting from its ordered phase  $h_{c1}^{(0)} < h < h_{c2}^{(0)}$
leads to a disorder-induced transition into a novel quantum disordered
phase. This all happens well away from the geometrical
percolation threshold $p^*$, which means that the disorder-induced
transition is a genuine quantum phase transition.

\begin{figure}[htb]
\centerline{\includegraphics[width=0.65\textwidth]{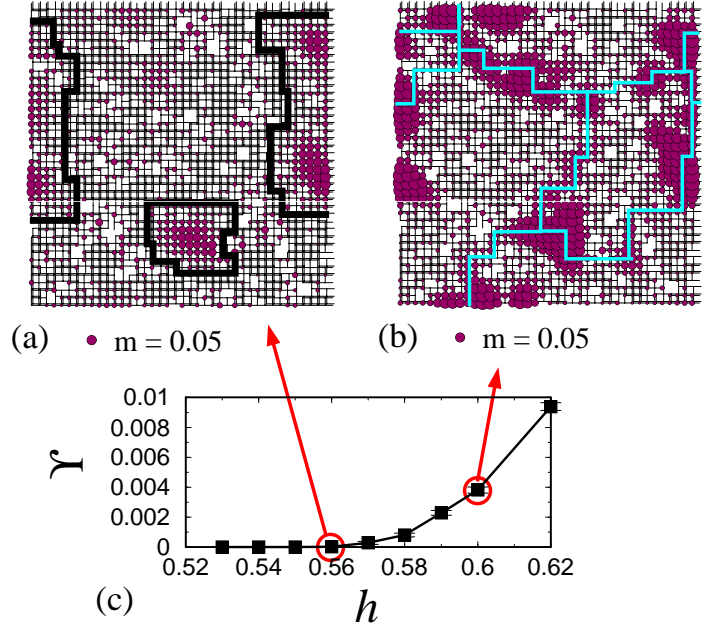}}
\caption{(a)-(b): Real-space images of the dimer magnetization
$m_i=\langle S^{z}_{i,1} + S^{z}_{i,2} \rangle$ on intact dimers in
a 40x40x2 bilayer with $J/J'=4$, dilution $p=0.1$ and at inverse
temperature $\beta J = 256$, for $h = 0.56$ (a) and $h = 0.6$ (b).
The radius of the dots is proportional to the dimer magnetization.
The magnetization of unpaired spins is omitted for clarity. The most
visible localized states are highlighted in (a), while the backbone
of the percolating magnetized network is highlighted in (b). (c)
Superfluid density (spin stiffness) as a function of the field for
the specific sample considered.} \label{f.LMBG}
\end{figure}

The novel phase that opens up at the disorder-induced
transition can be fully characterized in terms of bosonic
quasiparticles. In the case of  $h \gtrsim h_{c1}^{(0)}$
these quasi-particles (QPs) are represented by the dilute
triplet gas induced by the field, condensing in the
magnetically ordered phase. The case $h \lesssim h_{c2}^{(0)}$
is the (approximately) particle-hole symmetric one,
in which singlet quasi-holes (QHs) form a dilute gas that Bose
condenses. The introduction of site dilution in the system
creates an effective random potential for QPs
(analogous considerations apply to QHs).
Indeed site dilution leads to the disappearance of
whole dimers hosting a QP, or to the
appearance of dangling spins (local moments) which are
essentially all polarized
in the field range $h_{c1}^{(0)} < h < h_{c2}^{(0)}$,
and hence act as impenetrable barriers to QPs
(see below for
a detailed discussion of the physics of local moments).
Therefore the problem of the response of the magnetic system
to site dilution is analogous to that of the evolution of a Bose-Einstein
condensate upon increasing the strength of a random potential
in which the condensate is immersed. For a strong enough
random potential the bosonic system will be \emph{fragmented}
into Anderson-localized states, sitting in the rare regions which
are devoid of vacancies. This phase is
called \emph{Bose glass} in the literature of disordered interacting
bosons \cite{Fisheretal89}. Given that we are effectively
working in the grand-canonical ensemble, the transition
 from a condensed phase to a Bose glass is accompanied by a
 decrease in the population, due to the fact that disorder lowers
 dramatically the local chemical potential in the regions close
 to impurities. This same localization-condensation transition can
 be probed by varying the field at fixed disorder strength: in this
 case the transition is clearly understood as a reduction of the
 density of the QPs/QHs, induced by the decrease/increase
 of the magnetic field.
  When traversing this transition in the reversed
 sense, by driving it with either a field or with disorder, we observe
 that the region occupied by the QPs/QHs
 undergoes a quantum percolation transition from a localized
 disordered phase, in which only disconnected rare clean regions
 host QPs/QHs, to a percolated ordered
 phase, in which the regions hosting QPs/QHs
 connect to form a percolating network. This geometric
 transition can be directly visualized by calculating
 the local magnetization profile, as shown in Fig.~\ref{f.LMBG}.
The novel, percolative nature of this transition with respect
to the condensation in the clean case reveals itself
in the critical exponents, which can be determined numerically
via finite-size scaling \cite{Roscilde06,Yuetal08}.
The resulting values are quite different from those of the
transition in the clean system, and in particular it is found that
the dynamical critical exponent $z$ equals the
spatial dimension $d$, which is consistent with an early
theoretical prediction~\cite{Fisheretal89}.

\begin{figure}[htb]
\centerline{\includegraphics[width=0.65\textwidth]{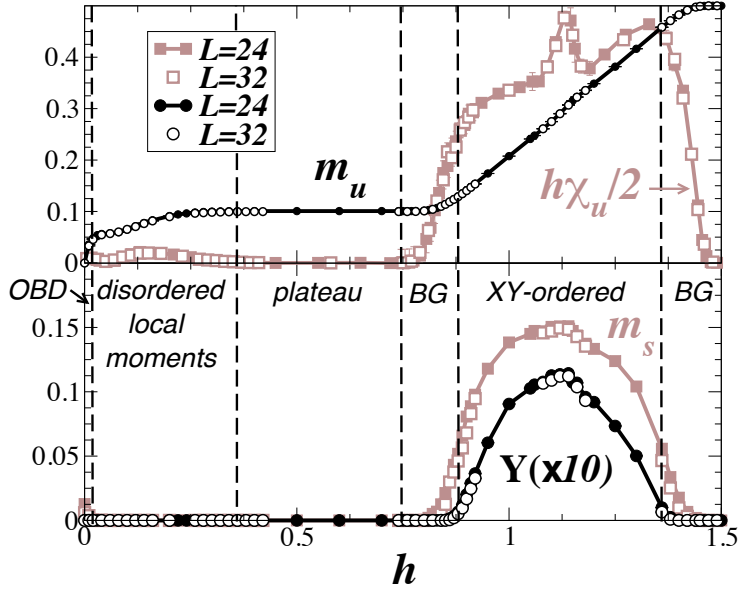}}
\caption{Zero-temperature field scan for the site-diluted Heisenberg
bilayer with $J/J'=8$ and $p=0.2$, from Roscilde \emph{et
al.},~\cite{Roscilde06}.} \label{f.FSBG}
\end{figure}

 From the point of view of macroscopic observables, the
 Bose glass is fundamentally dominated by the absence
 of a gap in its excitation spectrum. Indeed in this phase
 the system is fragmented into droplets of QPs/QHs
 that are hosted on rare, but arbitrarily large regions,
 and which can therefore support arbitrarily low-energy
 excitations, in the form of localized magnon excitations
 analogous to those appearing in the ordered phase in the
 clean system. This means that that the system displays
 a finite response to an applied magnetic field, namely
 its uniform susceptibility remains finite (corresponding
 to a finite compressibility of the QPs/QHs) even in absence
 of long-range coherence.  This is clearly illustrated
 in Fig.~\ref{f.FSBG}.

 So far we have worked under the assumption that
 dangling spins left unpaired by site dilution are fully
 polarized by the field. Actually the evolution of their
 magnetic state represents another very fascinating
 aspect of the physics of these systems. Indeed,
 as already mentioned in the previous sections,
 in zero field dilution liberates local $S=1/2$ moments (LMs)
 which are exponentially localized close to the site
 of an unpaired spin and which interact with
 each other via the effective couplings of Eq.~\ref{e.Jeff}.
 These couplings, although weak, are sufficient for the LMs to order
antiferromagnetically at experimentally relevant
temperatures~\cite{Azumaetal97}. Hence, at any finite dilution
concentration smaller than the classical percolation threshold, the
zero-field ground state displays long-range antiferromagnetic order
(order-by-disorder phase).

\begin{figure}[htb]
\centerline{\includegraphics[width=0.65\textwidth]{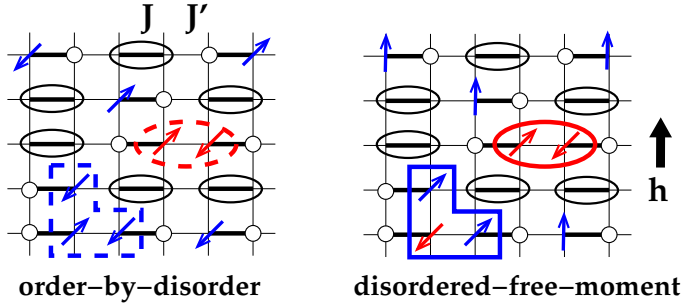}}
\caption{Sketch of the quantum phase transition between the order-by-disorder
phase and the disordered-local-moment phase in a
diluted coupled-dimer system.
\emph{Left panel}: At zero applied field, spins on intact dimers
form singlets (solid ellipses), dimers of local moments (LMs) have a
strong singlet component (dashed ellipse), whereas the other LMs
(blue arrows) participate in the order-by-disorder state. \emph{Right panel}: Upon
applying a field, the LMs are mostly polarized, but local
singlets and localized down-spins can survive on clustered LMs,
leading to the disordered LM phase.} \label{f.DLM}
\end{figure}

Yet the long-range order appearing in this phase can be
easily destroyed by a small field in a quite peculiar way
\cite{Roscilde06,Yuetal08}.
Indeed the system of LMs features highly inhomogeneous
couplings, and hence the response to the field is equally
diverse. In the case of low dilution (to which this discussion
is restricted) most of the LMs are weakly coupled to the
other nearby LMs which sit at a distance of the order
of the average inter-vacancy spacing $p^{-1/d}$; but
a minority of the LMs might be involved in a LM
dimer (with probability $\sim p^2$), in a trimer
  (with probability $\sim p^3$), and so on.
  This means that when the field is strong enough to
 overcome the \emph{typical} coupling energy between
 two LMs, it polarizes a majority of LMs destroying thereby
 their long-range magnetic order. Yet, in this
 \emph{disordered LM} phase, a minority of strongly
 coupled LMs resist polarization, and they host locally
 one or more spin which is polarized opposite to the field.
 In particular, a mapping of the localized $S=1/2$ spins which remain
 antiparallel to the field onto hardcore bosons, reveals for
 the disordered LM phase a clear nature of a Bose glass,
 characterized by fragmentation of a Bose gas into
 disconnected localized states.
This can be quantitatively confirmed by a finite-size scaling study of
the transition between the order-by-disorder phase and the disordered LM phase:
the extracted critical exponents turn out to be fully consistent with those of the
2D superfluid-to-Bose-glass transition \cite{Yuetal08}.
 In particular, as in a Bose glass, the disordered LM phase
 is not fully polarized, and it remains
 \emph{gapless}, given that all different clusters of strongly coupled LMs
 see a different magnetic environment and hence exhibit a
 different local gap to further polarization. Nonetheless
 the local gaps are distributed around some distinct
 values associated with LM dimers, trimers, etc.
 These values manifest themselves in a special form
 of the magnetization curve, exhibiting \emph{pseudo-plateaus}
 which mark the accomplished polarization of one class of
 LM clusters.

 \begin{figure}[htb]
\centerline{\includegraphics[width=0.65\textwidth]{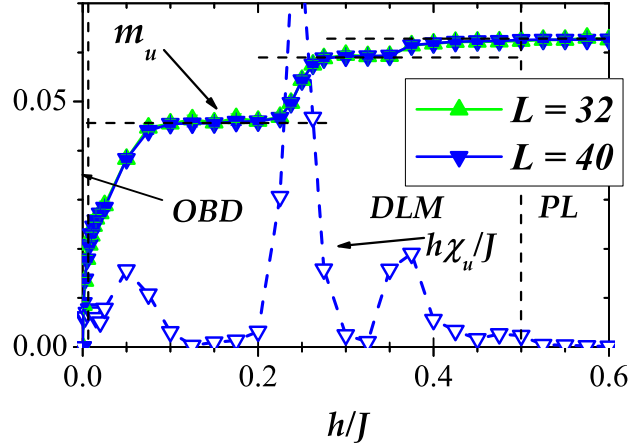}}
\caption{Magnetization and susceptibility curve of a planar
array of dimers with $J/J'=4$ and dilution $p=1/8$, showing
a characteristic succession of pseudo-plateaus.} \label{f.PP}
\end{figure}

 If the coupling ratio $J/J'$ is large enough, or the disorder
 concentration is weak enough, upon further increasing the field
 the LMs can be fully polarized at a field $h < h_{c1}^{(0)}$, namely
 before the magnetization process of the intact dimers begins.
 Under this assumption, a gapped \emph{plateau} phase
 (characterized by a marked plateau of the uniform magnetization
 at the value $m = p/2$) appears, completely separating the
 physics of the LMs from that of the intact dimers.
 On the other hand, for smaller $J/J'$ ratios or strong dilution
 the plateau phase might be completely washed out, and dimers
 start being magnetized in the Bose glass phase when LMs
 have not yet been fully polarized. In this case disordered
 LM phase and Bose glass phase coexist:
 indeed the Bose glass phase will show both localized
 triplet QPs in rare regions devoid of impurities, and
 localized anti-parallel spins in the complementary regions
 exhibiting clustering of impurities.

\section{Conclusions}

 In conclusion, large-scale quantum Monte Carlo simulations
give access to the vast wealth of novel quantum phases appearing
in quantum magnets when quantum fluctuations are tuned to a critical
strength in presence of disorder in the magnetic lattice.
These novel phases are the result of the separation, induced
by quantum fluctuations, between the
geometric percolation of the lattice and the loss of long-range
magnetic order. Their common denominator is a rich structure
of the low-energy spectrum, with absence of a gap and a possibly
anomalous role of the low-energy excitations in the response
functions.

 This rich physics enjoys the availability of a large family of
 $S=1/2$ compounds exhibiting strongly fluctuating quantum magnetism,
 and which are amenable to dilution of the magnetic lattice
 by non-magnetic doping.
 In particular candidate systems
 are spin-ladder materials such as  Cu$_{12}$(C$_5$H$_{12}$N$_2$)$_2$Cl$_4$
 ~\cite{Chaboussantetal97}
 and  (C$_5$H$_{12}$N$_2$)$_2$Cu Br$_4$ ~\cite{Watsonetal01},
 planar dimer systems such as Sr$_2$Cu(BO$_3$)$_2$ ~ \cite{Sebastianetal05-2},
 magnetic bilayer compounds as  BaCuSi$_2$O$_6$ ~ \cite{Sebastianetal05}.
 In all these compounds the $S=1/2$ spins is carried by a Cu$^{2+}$ ion,
 which can be replaced by non-magnetic Mg$^{2+}$ and Zn$^{2+}$
 to give site dilution of the lattice. Alternatively bond disorder
 can be realized by chemical substitution of the non-magnetic
 ions bridging the magnetic couplings~\cite{Oosawa02,Masuda04} .
 For what concerns specifically the physics of magnetic BEC in presence
 of disorder, analogous phenomena to those observed in $S=1/2$
 compounds can be recovered also with $S=1$ Haldane chains,
 or with  $S=1$ antiferromagnets with strong single-ion anisotropy
 \cite{Zapfetal06}.
We therefore believe that quantum magnets under static doping
offer the possibility of deepening tremendously our understanding
of quantum many-body physics in a controlled random environment.

\section{Acknowledgements}
We thank Omid Nohadani for useful discussion. This work was
supported by DOE through grant No. DE-FG02-06ER46319. The numerical
computations were carried out on the University of Southern
California high-performance computer cluster.


\begin{thebibliography}{99}

\bibitem{Sondhietal97} S. L. Sondhi, S. M. Girvin, J. P. Carini, and D. Shahar,
Rev. Mod. Phys. 1997, {\bf 69}, 315.
\bibitem{Giamarchi08} Giamarchi~T., R\"{u}egg~C., Tchernyshyov~O.,
Nat. Phys., 2008, \textbf{4}, 198.
\bibitem{Vajketal02} Vajk~O.P., {\it et al.}, Science, 2002, \textbf{295}, 1691.
\bibitem{Sandvik02} Sandvik~A.W., Phys. Rev. B, 2002,\textbf{66}, 024418.
\bibitem{VajkG02} Vajk~O.P., Greven~M.,
Phys. Rev. Lett., 2002, \textbf{89}, 177202.
\bibitem{Sandvik02B} Sandvik~A.W., Phys. Rev. Lett., 2002, \textbf{89},
177201.
\bibitem{Sknepneketal04} Sknepnek~R., Vojta~T., Vojta~M., Phys.
Rev. Lett., 2004, \textbf{93}, 097201.
 \bibitem{Yuetal05} Yu~R., Roscilde~T., Haas~S.,
        Phys. Rev. Lett., 2005, \textbf{94}, 197204.
\bibitem{RoscildeH05} Roscilde~T., Haas~S., Phys. Rev. Lett., 2005,
\textbf{95}, 207206.
\bibitem{RoscildeH06} Roscilde~T., Haas~S.,
J. Phys. B 2006, {\bf 39}, S153 (2006).

\bibitem{Roscilde06} Roscilde~T., Phys. Rev. B, 2006, \textbf{74},
144418.

\bibitem{Yuetal08} Yu~R., Roscilde~T., Haas~S., New J. Phys., 2008,
\textbf{10}, 013034.
\bibitem{LeeR85} Lee~P. A., and Ramakrishnan~T. V., Rev. Mod. Phys. 1985,
{\bf 57}, 287.
\bibitem{SigristF96} Sigrist~M. and Furusaki~A., J. Phys. Soc. Jpn. 1996,
{\bf 65}, 2385.
\bibitem{Mikeskaetal04} Mikeska~H.-J., Ghosh~A., and Kolezhuk~A. K.,
Phys. Rev. Lett. 2004, {\bf 93}, 217204.
\bibitem{Griffiths} Griffiths~R. B., Phys. Rev. Lett. 1969, {\bf 23}, 17;
McCoy~B. M., Phys. Rev. Lett. 1969, {\bf 23}, 383; McCoy~B. M.,
Phys. Rev. 1969, {\bf 188}, 1014.
\bibitem{Weichman08} Weichman~P. B., arXiv:0810.3263 2008.
\bibitem{SyljuasenSandvik02} Sylju{\aa}sen~O.F., Sandvik~A.W., Phys.
Rev. E, 2002, \textbf{66}, 046701.
\bibitem{Manousakis91} Manousakis~E., Rev. Mod. Phys., 1991,
    \textbf{63}, 1.
 \bibitem{StaufferA94} Stauffer~D., Aharony~A.,
        Introduction to Percolation Theory. Taylor and Francis, London,
        1994.
        \bibitem{SandvikS94} Sandvik~A. W. and Scalapino~D. J.,
Phys. Rev. Lett. 1994, {\bf 72}, 2777.
\bibitem{RoscildeH07} Roscilde~T., Haas~S., Phys. Rev. Lett., 2007,
\textbf{99}, 047205.
\bibitem{DagottoR96} E. Dagotto and T. M. Rice,
Science 1996, {\bf 271}, 618.
   \bibitem{Yuetal06} Yu~R., Roscilde~T., Haas~S.,
        Phys. Rev. B, 2006, \textbf{73}, 064406.
\bibitem{Matsumotoetal01} Matsumoto~M. \emph{et al.},
Phys. Rev. B 2001, {\bf 65}, 014407.

\bibitem{Fisheretal89} Fisher~M.P.A. \emph{et al.}, Phys. Rev. B, 1989,
\textbf{40}, 546.
\bibitem{Azumaetal97} Azuma~M. \emph{et al.}, Phys. Rev. B, 1997, \textbf{55}, R8658.
\bibitem{Chaboussantetal97} Chaboussant~G \emph{et al.}, Eur. J. Phys. B 1998, {\bf 6}, 167.
\bibitem{Watsonetal01} Watson~B. C. \emph{et al.}, Phys. Rev. Lett. 2001, {\bf 86}, 5168.
\bibitem{Sebastianetal05-2} S. E. Sebastian \emph{et al.}, Phys. Rev. B 2005, {\bf 71}, 212405.
\bibitem{Sebastianetal05} S. E. Sebastian \emph{et al.}, Phys. Rev. B 2005,
{\bf 72}, 100404.
\bibitem{Oosawa02} Oosawa~A., Tanaka~H., Phys. Rev. B, 2002,
\textbf{65}, 184437.
\bibitem{Masuda04} Masuda~T., Zheludev~A.,
Uchinokura~K., Chung~J.H., Park~S., Phys. Rev. Lett., 2004,
\textbf{93}, 077206.
\bibitem{Zapfetal06} A. Paduan-Filho \emph{et al.}, Phys. Rev. B 2004, {\bf
69}, 020405; V. Zapf \emph{et al.} 2006, Phys. Rev. Lett. {\bf 96},
077204.
















\end{thebibliography}
\end{document}